\documentclass[%
 aip,
 amsmath,amssymb,
reprint,%
]{revtex4-1}
\usepackage{graphicx}
\usepackage{dcolumn}
\usepackage{bm}

\usepackage[utf8]{inputenc}
\usepackage[T1]{fontenc}
\usepackage{mathptmx}
\usepackage{etoolbox}
\usepackage{soul,xcolor}
\setstcolor{red}

\renewcommand{\b}{\beta}

\newcommand{\add}[1]{\if\a\b{{\color{red} #1}}\else{#1}\fi}

\newcommand{\citeasnoun}[1]{Ref.~\onlinecite{#1}}

\renewcommand{\eqref}[1]{Eq.~(\ref{eq:#1})}

\newcommand{\figref}[1]{Fig.~\ref{fig:#1}}

\newcommand{\appref}[1]{Appendix~\ref{sec:#1}}

\makeatletter
\def\@email#1#2{%
 \endgroup
 \patchcmd{\titleblock@produce}
  {\frontmatter@RRAPformat}
  {\frontmatter@RRAPformat{\produce@RRAP{*#1\href{mailto:#2}{#2}}}\frontmatter@RRAPformat}
  {}{}
}%
\makeatother
\begin{document}
\title{Adaptive four-level modeling of laser cooling of solids}

\author{Weiliang Jin}
\affiliation{Department of Electrical Engineering, Ginzton Laboratory, Stanford University, Stanford, California 94305, USA}
\author{Cheng Guo}
\affiliation{Department of Applied Physics, Ginzton Laboratory, Stanford University, Stanford, California 94305, USA}
\author{Meir Orenstein}
\affiliation{Department of Electrical and Computer Engineering, Technion-Israel Institute of Technology, 32000 Haifa, Israel}
\author{Shanhui Fan}
\affiliation{Department of Electrical Engineering, Ginzton Laboratory, Stanford University, Stanford, California 94305, USA}
\email{shanhui@stanford.edu}
\date{\today}
\begin{abstract}
  Laser cooling of rare-earth doped solids has been demonstrated
  across a wide range of material platforms, inspiring the development
  of simple phenomenological models such as the four-level model to
  elucidate the universal properties of laser cooling under various
  operating conditions. However, these models usually require the
  input of full absorption spectra that must be provided
  experimentally or by additional complicated atomic modeling. In this
  letter, we propose that a four-level model, when extended to admit
  effective energy levels adaptive to the pumping photon energy, can
  accurately predict the cooling efficiency as a function of
  temperature and pumping frequency using only few inputs such as the
  absorption coefficient measured at a single frequency and
  temperature. Our model exploits the quasi-equilibrium properties of
  the excitation of rare-earth ions for the determination of the
  effective four energy levels. The model is validated against
  published experimental results for a span of materials including
  ytterbium/thulium-doped glass and crystals. With the verified model,
  we derive explicit expressions for the optimal frequency and the
  operating bandwidth of pumping laser. Our model significantly
  simplifies the modeling process of laser cooling, and is expected to
  stimulate further development of optical refrigeration.
\end{abstract}
\maketitle

Laser cooling of solids realizes refrigeration by using anti-Stokes
fluorescence with a fluorescence frequency higher than the pumping
frequency~\cite{seletskiy2012cryogenic,nemova2015twenty,HEHLEN2014179,nemova2016laser,ding2016active}. The
frequency difference corresponds to the phonon energy extracted from
the solids. Compared to traditional mechanical refrigeration, laser
cooling does not involve moving parts or cryogenic fluids, making it
advantageous for vibration-free applications and compact
cryocoolers. Intense research in recent decades has enabled laser
cooling in rare-earth doped fluoride crystals to achieve a remarkable
low temperature~\cite{melgaard2016solid} of $91$~K,
making optical refrigeration by far the only all-solid-state cryogenic
refrigeration technology.

\begin{figure}[htbp]
  \centering
  \includegraphics[width=0.46\textwidth]{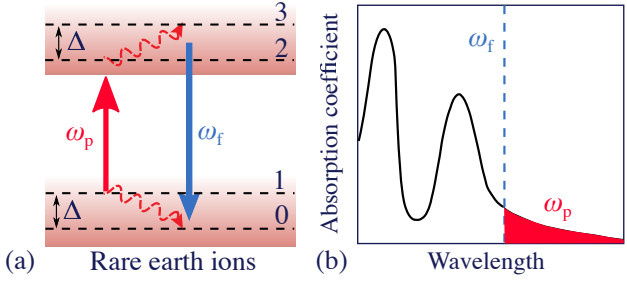}
  \caption{Schematic of anti-Stokes emission from the rare-earth ions
    whose mean fluorescence angular frequency $\omega_f$ is higher
    than the incident laser angular frequency $\omega_p$. (a) The
    participating energy manifolds (red shaded region) are
    approximated here by adaptive four energy levels (black dashed
    line). $\omega_{ij}\equiv\omega_i-\omega_j$ denotes the angular
    frequency separation between Level-$i$ and Level-$j$. The model
    assumes $\omega_{32}=\omega_{10}\equiv\Delta$. (b) A
    representative absorption spectrum illustrating that the pumping
    frequency $\omega_{21}=\omega_p$, lower than $\omega_f$ (blue
    dashed line) typically occurs in the absorption tail (red
    region).}
  \label{fig:scheme}
\end{figure}

The laser cooling process for a typical rare-earth ion involves
radiative transitions of near-unity quantum yield between the ground
and the excited manifolds.  As illustrated in \figref{scheme} (a),
each manifold contains multiple energy levels, thus a comprehensive
theory needs to capture all participating energy
levels~\cite{lamouche1998low,knall2018model,nemova2015optimization,nemova2016laser}
as well as their interactions with the phonons of the host
medium~\cite{fernandez2001origin,kim2008material}. To elucidate the
main physics and the impact of various material and operating
parameters, several simplified models such as quasi two-level
models~\cite{knall2018model,mobini2020laser} and four-level
models~\cite{seletskiy2012cryogenic,seletskiy2016laser} have been
developed. However, these models need extensive input of experimental
data such as the detailed fluorescence/absorption spectrum of the
rare-earth ions. It is therefore difficult to use these models to
reveal some of the general properties of laser cooling, such as the
general behaviors of many key figure of merits (FOM's). For instance,
a crucial FOM is the cooling efficiency $\eta_c$, defined as the net
cooling power normalized by the total pumping power. $\eta_c$ usually
exhibits complicated dependence on many parameters such as pumping
frequency and operating temperature.  In most laser cooling systems,
the optimal achievable cooling efficiency is usually assumed to
be~\cite{sheik2009laser,seletskiy2016laser}
$\eta_c\sim k_BT/\hbar\omega_p$ where $\omega_p$ is the pumping
angular frequency, $T$ the operating temperature, and $k_B$ the
Boltzmann constant. However, beyond a few qualitative arguments, we
are not aware of any quantitative formulation to support such a basic
result. As an another example, while it is known that laser cooling
only occurs within a narrow range of pumping frequencies and that this
bandwidth is temperature-dependent as observed
experimentally~\cite{seletskiy2012cryogenic,seletskiy2016laser}, a
quantitative description of temperature-dependence of such bandwidth
and potentially other factors that influence this bandwidth remain
unknown.

The lack of a simple semi-analytical model might be attributed to the
complex absorption spectra of rare-earth ions that contain many
resonance features at the cryogenic temperature ($<120$~K), which was
the main pursued application of laser
cooling~\cite{seletskiy2012cryogenic,nemova2016laser}. On the other
hand, at higher temperatures, as illustrated in \figref{scheme} (b),
the absorption spectrum in the cooling regime (red) tends to follow a
smooth exponential decay features, known as the cooling tail or the
long-wavelength tail~\cite{seletskiy2016laser}. A simple laser cooling
model developed in this domain can be beneficial for applications at
room-temperature or above~\cite{nakayama2019improving}, such as
cooling of light-driven
spacecraft~\cite{atwater2018materials,weiliang2021} and high-power
lasers where heat needs to be dissipated
efficiently~\cite{bowman1999lasers,knall2018model,balliu2019predictive,knall2020laser}.

In this letter, we develop a simple semi-analytical model to capture
the main features of laser cooling operating above the cryogenic
temperature. Our model, termed as the adaptive four-level model,
exploits the quasi-equilibrium properties of the excitation of
rare-earth ions~\cite{digonnet2001rare}, and extends the four-level
models~\cite{seletskiy2012cryogenic,seletskiy2016laser} to admit
effective energy levels that are determined by pumping photon
energy. Such a phenomenological model can accurately predict the
cooling efficiency over a wide range of temperatures and pumping
frequencies using only few inputs including the external quantum
yield, mean fluorescence frequency, parasitic absorption coefficient,
and absorption coefficient of the rare-earth ions measured at a single
value of temperature and frequency. The absorption spectrum in the
long-wavelength excitation regime evaluated from our model resembles
the Urbach tail~\cite{urbach1953long,toyozawa1959proposed}. We
validate our model against published experimental results in terms of
absorption coefficient and cooling efficiency over a wide range of
material choices including ytterbium-doped ZBLANP
glass~\cite{mungan1997laser,lei1998spectroscopic}, silica
glass~\cite{mobini2020laser}, YLF
crystal~\cite{seletskiy2011local,seletskiy2013precise}, and thulium
doped ZBLANP glass~\cite{hoyt2000observation}. With our model, we
prove that when maximizing the cooling power density in the ideal
case, the optimal pumping angular frequency $\omega_{p,\mathrm{opt}}$
needs to be redshifted from the mean fluorescence angular frequency
$\omega_f$ by $\omega_f-\omega_{p,\mathrm{opt}}\sim 0.8k_BT/\hbar$,
corresponding to a cooling efficiency
$\eta_c\sim 0.8 k_BT/\hbar\omega_p$. Additionally, we show that the
operating bandwidth, when normalized by the pumping frequency, is
approximately $\sim 1.2 k_BT/\hbar\omega_p$.

We begin by reviewing the prior four-level model of rare-earth doped
media in the context of laser
cooling~\cite{seletskiy2012cryogenic,seletskiy2016laser}. The
participating energy levels in the cooling process, as illustrated in
\figref{scheme} (a), lie in the two narrow manifolds called the
excited and the ground manifolds, each of which contains several
sublevels. The four-level
model~\cite{seletskiy2012cryogenic,seletskiy2016laser} approximates
each manifold as two energy levels. Under an incident laser beam of
angular frequency $\omega_p=\omega_{21}$ where
$\omega_{ij}=\omega_i-\omega_j$ denotes the angular frequency
separation between Level-$i$ and Level-$j$, electrons are excited from
the upper Level-$1$ of the ground manifold to the lower Level-$2$ of
the excited manifold, and then rapidly thermalized to the upper
Level-$3$ within the excited manifold, resulting in spontaneous
emission of photons of higher frequencies.

The cooling performance can be estimated from the steady-state
solution of the electron occupation distribution over the four energy
levels. Typically, this is done by applying the standard rate equation
approach~\cite{siegman1986lasers,seletskiy2012cryogenic}. In what
follows, we make the typical simplification by assuming equal
inter-manifold transition rate, denoted as $\gamma/2$, so that the
lifetime of the Level-$2(3)$ is $\gamma^{-1}$. $\gamma$ is orders of
magnitude lower than the intra-manifold transition rate. Additionally,
we assume equal width of the two manifolds, $\omega_{32}=\omega_{10}$,
denoted as $\Delta$. Considering a pumping intensity $I$, the absorbed
power $P_{\mathrm{abs}}=[\alpha_b+\alpha]I$ is attributed to the
parasitic (background) absorption $\alpha_b$ and rare-earth ion
absorption $\alpha$. Two important FOMs, the net cooling power density
$P_{\mathrm{cool}}$ and the cooling efficiency
$\eta_c=\frac{P_{\mathrm{cool}}}{P_{\mathrm{abs}}}$, are given by (The
details of the derivation can be found in \appref{rate}),
\begin{align}
  &P_{\mathrm{cool}} = \left[ \eta_e \frac{\omega_f}{\omega_p}-1 \right]\alpha I-\alpha_bI\label{eq:cool}\\
  &\eta_c=\eta_e\eta_{\mathrm{abs}} \frac{\omega_f}{\omega_p}-1\label{eq:eta}
\end{align}
Here, $\omega_f$ is the mean fluorescence angular frequency,
$\eta_e=\gamma_r/\gamma$ the external quantum yield defined as the
ratio between the radiative transition rate $\gamma_r$ and the total
transition rate $\gamma$, and
$\eta_{\mathrm{abs}}=\frac{\alpha}{\alpha+\alpha_b}$ the absorption
efficiency. Eqs.~(1-2) are well-known, and they can also be obtained
with other approaches such as the quasi two-level
model~\cite{knall2018model,mobini2020laser}. Among several parameters
in Eqs. (1-2), the absorption coefficient $\alpha$ is known to exhibit
strong dependence on both $\omega_p$ and temperature. Thus, the
magnitude of the two FOMs are largely dictated by
$\alpha(\omega_p,T)$. As detailed in \appref{rate}, the four-level
model provides a solution for $\omega_f$ and $\alpha$,
\begin{align}
  &\omega_f = \omega_{p} + \Delta \left[ \frac{1}{2}+\frac{1}{1+\exp[\hbar\Delta/k_BT]} \right]\label{eq:mean}\\
  &\alpha = \frac{\sigma_{12}N_t}{1+\exp[\hbar\Delta/k_BT]}\frac{1}{1+I/I_s}\label{eq:alpha}
\end{align}
where $\sigma_{12}$ is the atomic absorption cross section for the
transition $1\rightarrow2$, $N_t$ the doping concentration, and
$I_s=\hbar\omega\gamma/\sigma_{12}$ the saturation intensity.
The exponential factor $\exp[\hbar\Delta/k_BT]$ is attributed to the
Boltzmann distribution of the electron occupation
density~\cite{siegman1986lasers} across distinguishable localized
rare-earth ions. Results in Eqs. (3-4) cannot be obtained with the
quasi two-level model. They provide physical
insights~\cite{seletskiy2012cryogenic,seletskiy2016laser} into the
temperature-dependence of the cooling performance.

However, the application of the former four-level model is limited in
practice. It assumes that the manifold width $\Delta$ of the
four-level system is specified by the manifold width of the actual
ion, which has no relation to the incident laser frequency
$\omega_p$. In this case, \eqref{alpha} does not reveal useful
spectroscopic information of $\alpha$ as a function of
$\omega_p$. Consequently, to capture $\eta_c$ over a range of
$\omega_p$, the experimentally measured absorption spectrum
$\alpha(\omega_p)$ needs to be employed in place of
\eqref{alpha}. Another modification of the four-level model is to use
a phenomenological $\alpha(\omega_p)$ by incorporating the linewidth
broadening effect~\cite{seletskiy2012cryogenic,seletskiy2016laser}
into the atomic absorption cross section $\sigma_{12}$. But the
determination of the linewidth function, which is the superposition of
many Lorentzian and Gaussian lineshape
functions~\cite{martin2006reciprocity}, again undermines the
simplicity of the results as derived from the four-level model.


We address these limitations through better approximating the electron
occupation distribution by assigning effective participating energy
levels whose energy values are solved for at each $\omega_p$, so that
the energy levels will adapt to $\omega_p$. The levels are determined
by exploiting the quasi-equilibrium property of the electrons within
each manifold of the rare-earth ions~\cite{digonnet2001rare}, which
will result in a $\omega_p$-independent mean fluorescence frequency
$\omega_f$. This is also known as Vavilov's law based on empirical
observations~\cite{digonnet2001rare,rayner2003condensed}. Once
$\omega_f$ is provided, for each value of $\omega_p$, \eqref{mean} is
employed to solve for $\Delta(\omega_p)$, which together with
$\omega_{21}=\omega_p$ fully determines the relative positions of the
four levels. Thus, we call this model ``adaptive four-level
model''. For a laser cooling scheme, the solution $\Delta(\omega)$ is
subsequently used to obtain the spectrum $\alpha(\omega_p)$ via
\eqref{alpha}, and finally the cooling FOMs Eqs. (1-2).
As for the number of the input parameters, in the unsaturated regime
where $I\ll I_s$, only three ($\sigma_{12}$, $N_t$ and $\omega_f$) are
needed to determine $\alpha(\omega_p)$, and two more ($\eta_e$ and
$\alpha_b$) for $\eta_c$. In practice, all these five parameters can
be assumed to be $\omega_p$-independent. The assumption of an
undispersive $\sigma_{12}$ is in particular applicable to high
temperature (e.g. above the cryogenic temperature) and disordered
solids such as glass. The five parameters are also roughly
temperature-insensitive except for $\omega_f$ that can exhibit a weak
linear temperature
dependence~\cite{seletskiy2013precise,seletskiy2012cryogenic}. Therefore,
compared to the former four-level model that requires the input of
$\alpha(\omega_p,T)$ over all relevant pumping frequencies and
temperatures, our model can present great simplification, as well as
more transparent description of the temperature- and
$\omega_p$-dependence of the cooling performance.

To gain insights, we further study the model in the high phonon energy
extraction region $\hbar\Delta\gg k_BT$ where Eqs. (3-4) can be
simplified. The manifold width is $\Delta\approx2(\omega_f-\omega_p)$,
and the absorption exhibits the expected exponential decay feature,
\begin{equation}
\alpha(\omega_p) \approx \frac{\sigma_{12}N_t}{1+I/I_s}\exp[-2\hbar(\omega_f-\omega_p)/k_BT]\label{eq:alphal}
\end{equation}
Below we make a few observations. First, the solution $\Delta$ can
exceed the actual manifold width of the rare-earth ion, indicating
that Level-$1$ and Level-$2$ can lie in the phonon-broadened
sidebands.
Second, the exponential decay feature in \eqref{alphal} resembles the
Urbach tail in insulators and
semiconductors~\cite{urbach1953long,toyozawa1959proposed} arising from
the electron-phonon interactions driven by thermal fluctuations. The
absorption coefficient in the Urbach tail is proportional to
$\exp[-\sigma\hbar(\Omega-\omega_p)/k_BT]$ where $\hbar\Omega$ is the
effective energy gap, and $\sigma$ a constant of the order of one. Our
phenomenological model corresponds to a Urbach tail of
$\Omega=\omega_f$ and $\sigma=2$. Such a similarity can be attributed
to the thermalization process present in both cases. In fact, the
modeling of the optical transitions in semiconductors for
photoluminescence (PL)
applications~\cite{wurfel1982chemical,cardona2005fundamentals} is also
similar to our model except for modifications such as the use of
chemical potential in place of rate equations, and Fermi distribution
in place of Boltzmann distribution.


\begin{figure}[htbp]
  \centering
  \includegraphics[width=0.48\textwidth]{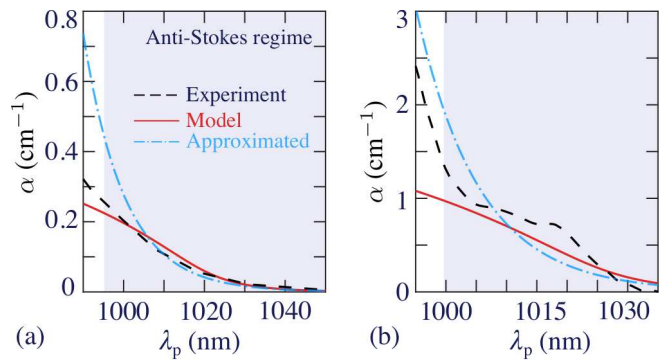}
  \caption{The absorption spectrum at room temperature evaluated from
    the adaptive four-level model (red solid line), the approximated
    model described by \eqref{alphal} (blue dash-dotted line), in
    comparison to the experimental results (black dashed line) of (a)
    Yb$^{3+}$ doped ZBLANP glass~\cite{lei1998spectroscopic} with
    doping concentration $1.128\times10^{20}$~cm$^{-3}$ and (b)
    Yb$^{3+}$ doped YLF crystal~\cite{seletskiy2013precise} with
    doping concentration $6.98\times10^{20}$~cm$^{-3}$. The only
    fitting parameter in the model is $\sigma_{12}$. The three agree
    well in the anti-Stokes emission regime (shaded region), where the
    pumping wavelength is redshifted from the mean fluorescence
    wavelength.}
  \label{fig:abs}
\end{figure}

We validate our model against experimental results in terms of
absorption spectrum and cooling efficiency. \figref{abs} depicts the
room-temperature absorption spectrum evaluated from the adaptive
four-level model (red solid line), the approximated model in the high
phonon energy extraction region (blue dash-dotted line), and
experimental results (black dashed line). For the sake of generality,
we cover two main types of the host media, glass and crystals,
represented by ZBLANP glass in \figref{abs} (a), and YLF crystal in
\figref{abs} (b), respectively. The rare-earth ion is taken as
Yb$^{3+}$ due to its dominance in the literature and hence the
accessibility of published experimental data. The computation of the
spectrum with our model involves only one fitting parameter
$\sigma_{12}$, while the other two parameters $\omega_f$ and $N_{t}$
are obtained from the experimental data. The detailed values are
listed in the \appref{parameter}.

\begin{figure*}[htbp]
  \centering
  \includegraphics[width=1\textwidth]{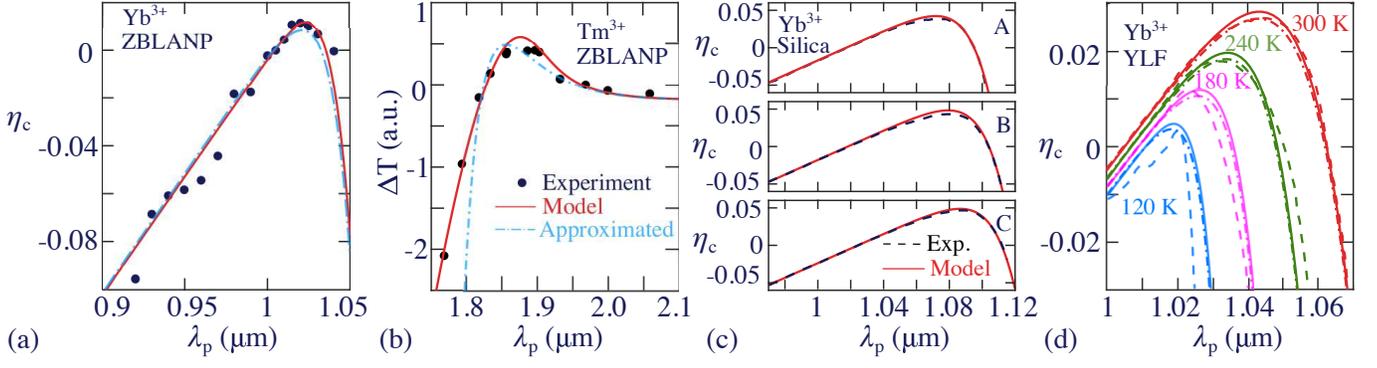}
  \caption{The cooling efficiency $\eta_c$ or temperature change
    $\Delta T$ (proportional to $\eta_c$) as a function of pumping
    wavelength based on the adaptive four-level model (solid line),
    the approximated model (dash-dotted line), in comparison to the
    experimental results (black dots in (a)(b) and dashed lines in
    (c)(d)). Cooling occurs when $\eta_c>0$ or $\Delta T>0$. The
    material platforms are (a) Yb$^{3+}$ doped ZBLANP
    glass~\cite{mungan1997laser}, (b) Tm$^{3+}$ doped ZBLANP
    glass~\cite{hoyt2000observation}, (c) three samples of Yb$^{3+}$
    doped silica glass prepared with different fabrication
    techniques~\cite{mobini2020laser} denoted as $A$, $B$, and $C$,
    and (d) Yb$^{3+}$ doped YLF crystal~\cite{seletskiy2011local}
    operating at various temperatures. The solid and the dashed-dotted
    curves in (c) coincide. The experimental results in (a)(b) are
    inferred directly from the temperature change measurement, while
    those in (c)(d) are derived from the absorption/fluorescence
    spectrum.}
  \label{fig:validate}
\end{figure*}

\figref{abs} (a) shows that for Yb$^{3+}$ doped ZBLANP glass, the
measured absorption spectrum (black dashed line) indeed exhibits a
smooth exponential decay lineshape in the anti-Stokes emission regime
where the pumping wavelength $\lambda_p$ is redshifted from the mean
fluorescence wavelength $\lambda_f$. The spectrum computed from the
adaptive four-level model agrees well with the experiment in almost
the entire anti-Stokes emission regime. The approximate model of
\eqref{alphal} also exhibits good agreement when the photon energy of
the incidence is redshifted from that of the fluorescence by at least
one $k_BT$.
As for Yb$^{3+}$ doped YLF crystal, as shown in \figref{abs} (b), the
measured absorption spectrum in the anti-Stokes emission regime
exhibits additional features on top of the exponential decay
tail. This is because the inhomogeneous broadening, which dominates in
glass, is suppressed in a crystalline media, allowing the resonance
absorption peaks to be enhanced in a crystal. However, those features
occurring in the long-wavelength tail are still largely suppressed
since the electron occupation density on high energy levels are low
according to the Boltzmann distribution. Indeed, \figref{abs} (b)
elucidates that our model can still capture the general trend of the
spectrum. The impact of the missing little ``bumps'' in our model on
the cooling efficiency will be studied below.

\begin{figure}[htbp]
  \centering
  \includegraphics[width=0.48\textwidth]{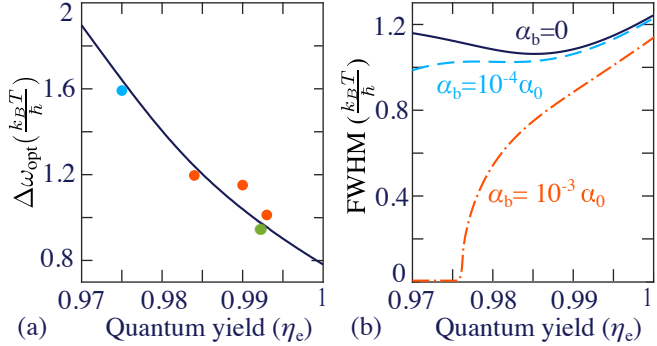}
  \caption{(a) The optimal frequency detuning
    $\Delta\omega_{\mathrm{opt}}$ of the pumping light from the mean
    fluorescence frequency for maximizing cooling power density as a
    function of external quantum yield evaluated from the adaptive
    four-level model (black solid line) or experimental
    results~\cite{mobini2020laser,knall2020laser,mungan1997laser}
    (dots). Dots with different colors are taken from different
    experimental reports, detailed in \appref{parameter}. (b) The
    spectral full width at half maximum of the cooling power density
    as a function of external quantum yield at various values of
    parasitic absorption rate. $\alpha_0$ denotes the absorption
    coefficient of the ion at the mean fluorescence frequency.}
  \label{fig:eta}
\end{figure}

Next, we analyze the applicability of our model in capturing cooling
efficiency $\eta_c$. In \figref{validate}, we apply our model to a
wide range of material platforms in comparison to the experimental
results. In particular, among the three rare-earth ions (Yb$^{3+}$,
Tm$^{3+}$, and Er$^{3+}$) that have so far exhibited bulk
cooling~\cite{HEHLEN2014179}, \figref{validate} covers the results for
both Yb$^{3+}$ and Tm$^{3+}$; as for the host media, the recently
explored silica glass~\cite{mobini2020laser} is also discussed. In
evaluating $\eta_c$, only one parameter $\sigma_{12}$ needs to be
fitted~\footnote{When the values of $\alpha_b$ and $\eta_e$ are
  missing in the experimental reference, we obtain these values by
  fitting.}. \figref{validate} (a-c) summarize the room-temperature
cooling efficiency as a function of pumping wavelength based on our
model compared to experimental results for Yb$^{3+}$ doped ZBLANP
glass [\figref{validate} (a)], Tm$^{3+}$ doped ZBLANP glass
[\figref{validate} (b)], and Yb$^{3+}$ doped silica glass
[\figref{validate} (c)].  Remarkably, in all scenarios, the adaptive
four-level model agrees well with the experiments. The approximated
model performs well in \figref{validate} (a,c) involving Yb$^{3+}$
ions, but deviates from the experiment at shorter wavelengths in
\figref{validate} (b) related to Tm$^{3+}$ ions. This is because the
operating frequency of Tm$^{3+}$ ions is around half compared to
Yb$^{3+}$ ions, making the valid regime of the high phonon energy
extraction region defined in terms of $k_BT$ to shift to a relatively
longer wavelength.

In \figref{validate} (d), we further examine the accuracy of our model
in revealing the temperature-dependence of $\eta_c$. Except for
$\omega_f(T)$ that needs an experimental input, the impact of
operating temperature in our model is explicit in Eqs. (3-4), or
\eqref{alphal} in the long-wavelength excitation
regime. \figref{validate} (d) shows that when the operating
temperature is above $120$~K, the prediction of our model is very
accurate. At temperature $120$~K (blue curves) where our model is
becoming less appropriate, we capture the main trend, but do not
follow some fine features such as the two small peaks and slightly
over-estimate the cooling efficiency. As pointed out earlier, the
absorption spectrum at room temperature, as shown in \figref{abs} (b),
also exhibits small features that have negligible effects on the
cooling efficiency. Consequently, our model is applicable also for
crystalline host media.


Finally, we reveal several universal properties of laser cooling based
on our adaptive four-level model. First, in a laser cooling system, it
is crucial to know the optimal pumping frequency
$\omega_{p,\mathrm{opt}}$ that maximizes $P_{\mathrm{cool}}$, and the
corresponding value of $\eta_c$ at
$\omega_{p,\mathrm{opt}}$. \eqref{cool} appears to indicate that a
lower $\omega_p$ enhances $P_{\mathrm{cool}}$ by increasing the
extracted phonon energy. However, in practice the optimal
$\hbar\omega_p$ is only below $\hbar\omega_f$ by several thermal
energy $k_BT$ due to the diminishing of the absorption rate
$\alpha(\omega_p)$ with frequency. In the ideal scenarios where
$\eta_e\approx1$ and $\alpha_b\approx0$, our model shows that the peak
cooling power density is achieved at an angular frequency detuning of
$\Delta\omega_{p,\mathrm{opt}}\equiv\omega_f-\omega_{p,\mathrm{opt}}\approx0.78
\frac{k_BT}{\hbar}$, where the cooling efficiency yields,
\begin{equation}
\eta_c\approx0.78 \frac{k_BT}{\hbar\omega_p}\label{eq:peaketa}
\end{equation}
Our results provide a quantitative support to the previous
claims~\cite{seletskiy2016laser} that the optimal detuning should be
$\sim k_BT/\hbar$ and the optimal $\eta_c\sim k_BT/\hbar\omega_p$. As
shown in \figref{eta} (a), we also validate our prediction of
$\omega_{p,\mathrm{opt}}$ (black line) under non-ideal $\eta_e$
against several experimental results (dots, detailed in
\appref{parameter}). In general, at lower $\eta_e$, larger frequency
detuning to increase the extracted phonon energy is needed to
compensate for the heating due to the nonradiative process.

The second important quantity of laser cooling is the operating
bandwidth of the pumping frequency for achieving net refrigeration,
which in practice is relevant in choosing an appropriate laser. The
full width at the half maximum (FWHM) of the spectral peak of
$P_{\mathrm{cool}}$ can be determined in the ideal scenario of our
model as,
\begin{equation}
\frac{\mathrm{FWHM}}{\omega_p}\approx1.24 \frac{k_BT}{\hbar\omega_p}\label{eq:FWHM}
\end{equation}
\eqref{FWHM} reveals that there is a direct connection between the
optimal $\eta_c$ described by \eqref{peaketa} and the relative
operating bandwidth of the pumping frequency. Therefore, previous
efforts in enhancing $\eta_c$ such as operating at higher
temperature~\cite{nakayama2019improving} or exploring rare-earth ions
of lower transition frequencies (e.g. compared to Yb$^{3+}$ that
operates at $\sim1$~$\mu$m, Tm$^{3+}$ is at $\sim2$~$\mu$m) can also
be applied to increase the operating bandwidth. The operating
bandwidth in the non-ideal scenarios is depicted in \figref{eta} (b),
showing that while the FWHM is in general on the order $k_BT/\hbar$,
it can vanish rapidly with degrading $\eta_e$ when the parasitic
absorption is significant.

To conclude, we have proposed a simple phenomenological model of very
few inputs for capturing the spectral and thermal properties of laser
cooling. Our model is validated against experimental results across
several representative material platforms. Our model is mostly
accurate in the presence of substantial homogeneous or inhomogeneous
broadening effects where the lineshape of the absorption spectrum is
mostly determined by the electron thermalization process. A potential
simple model beyond this region (e.g. crystalline host media of small
inhomogeneous broadening held at the cryogenic temperature for
suppressing homogeneous broadening) is expected in the future.

\begin{acknowledgements}
  We would like to thank Dr. Wei Li and Dr. Alex Song for useful
  discussions. This work is supported by the Breakthrough Starshot
  Initiative (initial idea and performance testing), and the
  U.S. Department of Energy ``Photonics at Thermodynamic Limits''
  Energy Frontier Research Center under Grant No. DE-SC0019140
  (theoretical modeling).
\end{acknowledgements}

\section*{Data Availability Statement}
The data that support the findings of this study are available from
the corresponding author upon reasonable request.

\appendix

\section{Four-level model\label{sec:rate}}
\begin{table}[htbp]
  \centering
  \caption{Values of the parameters used to evaluate the absorption coefficient in Fig. 2 of the main text.}  
  \begin{tabular}{lccc}
    \hline\hline
    &$\lambda_f$ (nm)& $N_t$ ($10^{20}$ cm$^{-3}$)& $\sigma_{12}$ ($10^{-20}$ cm$^{-2}$)\\\hline
    ZBLANP glass~\cite{lei1998spectroscopic} & 995 (\citeasnoun{lei1998spectroscopic}) &1.128 (\citeasnoun{lei1998spectroscopic}) &0.4 (fitted) \\
    YLF crystal~\cite{seletskiy2013precise} & 999.6 (\citeasnoun{seletskiy2013precise}) &6.98 (\citeasnoun{seletskiy2013precise}) &0.28 (fitted)\\\hline\hline
  \end{tabular}
  \label{tab:abs}
\end{table}

\begin{table*}[htbp]
  \centering
  \caption{Values of the parameters used to evaluate the cooling
    efficiency in Fig. 3 of the main text.}
  \begin{tabular}{llllll}
    \hline\hline
    &$\lambda_f$ (nm) & $N_t$ ($10^{20}$ cm$^{-3}$) &$\eta_e$ &$\alpha_b$ ($10^{-4}$cm$^{-1}$)  & $\sigma_{12}$ ($10^{-20}$ cm$^{-2}$)\\\hline
    Yb$^{3+}$ -- ZBLANP glass~\cite{mungan1997laser}& 995 (\citeasnoun{mungan1997laser})  &2.42 (\citeasnoun{mungan1997laser}) &0.992 (fitted) &$0.8$ (fitted) &0.037 (fitted) \\
    Yb$^{3+}$ -- YLF crystal~\cite{seletskiy2011local}  &1008.9-0.031T (\citeasnoun{seletskiy2013precise}) &6.98 (\citeasnoun{seletskiy2011local}) &0.995 (\citeasnoun{seletskiy2011local}) &4 (\citeasnoun{seletskiy2011local}) & 0.28 (fitted)\\
    Yb$^{3+}$ -- silica glass A~\cite{mobini2020laser} &1010 (\citeasnoun{mobini2020laser}) &0.53 (\citeasnoun{mobini2020laser})&0.993 (\citeasnoun{mobini2020laser})&0.23 (\citeasnoun{mobini2020laser}) &0.9 (fitted)\\
    Yb$^{3+}$ -- silica glass B~\cite{mobini2020laser}  &1008 (\citeasnoun{mobini2020laser})&0.44 (\citeasnoun{mobini2020laser})&0.990 (\citeasnoun{mobini2020laser})&0.21 (\citeasnoun{mobini2020laser})&2.2 (fitted)\\
    Yb$^{3+}$ -- silica glass C~\cite{mobini2020laser}  &1008 (\citeasnoun{mobini2020laser})&0.57 (\citeasnoun{mobini2020laser})&0.984 (\citeasnoun{mobini2020laser})&0.12 (\citeasnoun{mobini2020laser})&1.7 (fitted)\\
    Tm$^{3+}$ -- ZBLANP glass~\cite{hoyt2000observation}   &1820 (\citeasnoun{hoyt2000observation}) &2.42 (\citeasnoun{hoyt2000observation})&0.9997 (fitted)&2 (fitted)&0.042 (fitted)\\\hline\hline
  \end{tabular}
  \label{tab:cool}
\end{table*}

\begin{table*}[htbp]
  \centering
  \caption{Values of the parameters used to evaluate the data points
    of the optimal pumping frequency detuning
    $\Delta\omega_{\mathrm{opt}}=\omega_f-\omega_{p,\mathrm{opt}}$ as
    a function of $\eta_e$ in Fig. 4 (a) of the main text, where
    $\omega_{p,\mathrm{opt}}$ is the pumping frequency that optimizes
    the cooling power.}
  \begin{tabular}{lccc}
    \hline\hline
    &$\eta_e$ &$\lambda_f$ (nm) &$\lambda_{p,\mathrm{opt}}$ (nm)\\\hline
    Yb$^{3+}$ doped silica glass A~\cite{mobini2020laser} (red dot) &0.993 (\citeasnoun{mobini2020laser})&1010 (\citeasnoun{mobini2020laser})&1032 (Read from the figure in \citeasnoun{mobini2020laser})\\
    Yb$^{3+}$ doped silica glass B~\cite{mobini2020laser} (red dot) &0.990 (\citeasnoun{mobini2020laser})&1008 (\citeasnoun{mobini2020laser})&1033 (Read from the figure in \citeasnoun{mobini2020laser})\\
    Yb$^{3+}$ doped silica glass C~\cite{mobini2020laser} (red dot) &0.984 (\citeasnoun{mobini2020laser})&1008 (\citeasnoun{mobini2020laser})&1034 (Read from the figure in \citeasnoun{mobini2020laser})\\
    Yb$^{3+}$ doped silica glass~\cite{knall2020laser} (blue dot) &0.975 (\citeasnoun{knall2020laser}) &1002.5 (\citeasnoun{knall2020laser})&1037 (\citeasnoun{knall2020laser})\\
    Yb$^{3+}$ doped ZBLANP glass~\cite{mungan1997laser} (green dot) &0.992 (fitted) &995 (\citeasnoun{mungan1997laser}) &1015 (\citeasnoun{mungan1997laser})\\
    \hline\hline
  \end{tabular}
  \label{tab:eta}
\end{table*}

Below we briefly review the rate equation formulation of laser cooling
based on four-level ions~\cite{seletskiy2012cryogenic}. As
schematically illustrated in \figref{scheme} of the main text, each of
the manifolds of the ground state and the excited state is
approximated as two closely-spaced energy levels. As a simple model,
we assume equal width $\Delta$ for both the ground and the excited
manifolds. We also assume equal degeneracy of all levels, and equal
decay rate for transitions between all the excited and the ground
states, namely,
$\gamma_{30}=\gamma_{31}=\gamma_{21}=\gamma_{20}\equiv
\frac{\gamma}{2}$. In the presence of laser pumping of frequency
$\omega_{p} = \omega_{21}$ between Level-$1$ and $2$, the electron
occupation density of each energy level evolves according to the
following rate equation,
\begin{align}
  &\frac{\mathrm{d}N_0}{\mathrm{d}t}= \frac{\gamma}{2}(N_2+N_3)+\gamma_{10}(N_1-N_0\exp[-\hbar\Delta/k_BT])\nonumber\\
  &\frac{\mathrm{d}N_2}{\mathrm{d}t}=\mathcal{P}(N_1-N_2)-\gamma N_2+\gamma_{32}(N_3-N_2\exp[-\hbar\Delta/k_BT])\nonumber\\
  &\frac{\mathrm{d}N_3}{\mathrm{d}t}=-\gamma N_3-\gamma_{32}(N_3-N_2\exp[-\hbar\Delta/k_BT])\nonumber\\
  &N_0+N_1+N_2+N_3 = N_t
\end{align}
where $N_t$ is the total electron density, and
$\mathcal{P}=\frac{I\sigma_{12}}{\hbar\omega_p}$ is the pumping rate
that depends on the laser intensity $I$ and ion absorption cross
section $\sigma_{12}$ of the Level $1\rightarrow2$ transition. For
rare-earth ions, we apply the approximate
$\gamma\ll \gamma_{32}(\gamma_{10})$ since the lifetime $1/\gamma$ is
typically on the order of $10^{-6}\sim10^{-3}$~s, while
$1/\gamma_{32}(\gamma_{10})$ is around $10^{-12}\sim10^{-9}$~s, as
seen in \citeasnoun{HEHLEN2014179}. The stationary solution relevant
to laser cooling yields,
\begin{align}
  &N_1-N_2=\frac{N_t}{(1+\frac{\mathcal{P}}{\gamma})(1+\exp[\hbar\Delta/k_BT])}\nonumber\\
  &N_2+N_3=\frac{\mathcal{P}}{\gamma}(N_1-N_2)\nonumber\\
  &N_2 =  \frac{\mathcal{P}}{\gamma}\frac{N_1-N_2}{(1+\exp[-\hbar\Delta/k_BT])}
\end{align}

With the foregoing expressions, the absorbed power and emitted power
can be expressed as,
\begin{align}
  &P_{\mathrm{abs}} = \mathcal{P}\hbar\omega(N_1-N_2)=\alpha I\\
  &P_{\mathrm{emit}} = (N_2+N_3)\hbar\omega_f \gamma_r = \eta_e \frac{\omega_f}{\omega_p}\alpha I
\end{align}
where $\gamma_r$ is the radiative decay rate, and the expression of
$\alpha$ can be found in the main text. The mean fluorescence
frequency $\omega_f$ in our four-level system is the average
transition frequency weighted by the occupation density,
\begin{align}
  \omega_f&=\frac{1}{2} \left[ \frac{N_2}{N_2+N_3}(\omega_{21}+\omega_{20})+\frac{N_3}{N_2+N_3}(\omega_{31}+\omega_{30}) \right]\nonumber\\
  & = \omega_p + \Delta \left[ \frac{1}{2}+\frac{1}{1+\exp[\hbar\Delta/k_BT]} \right]
\end{align}

\section{Parameter values\label{sec:parameter}}
Table (I-III) list parameter values for Fig.2, Fig.3, and Fig.4 (a),
respectively.


\begin{thebibliography}{34}
\bibitem [{\citenamefont {Jin}\ and\ \citenamefont {et~al}()}]{weiliang2021}%
  \BibitemOpen
  \bibfield  {author} {\bibinfo {author} {\bibfnamefont {W.}~\bibnamefont
  {Jin}}\ and\ \bibinfo {author} {\bibnamefont {et~al}},\ }\href@noop {}
  {\bibinfo  {journal} {To be published}\ }\BibitemShut {NoStop}%
\end{thebibliography}
\end{document}